\documentclass[conference]{IEEEtran}
\usepackage{cite}

\ifCLASSINFOpdf
  \usepackage[pdftex]{graphicx}
\else
\fi
\usepackage{url}

\usepackage{fixltx2e}

\begin{document}
%
\title{Analysis of Chromosome 20 - A Study}

\author{\IEEEauthorblockN{Kristiina Ausmees,Pushpam Aji John}
\IEEEauthorblockA{Department of Information Technology\\
Uppsala University, Sweden\\
 }} 


%


\maketitle

\begin{abstract}

Since the arrival of next-generation sequencing technologies the amount of genetic sequencing data has increased dramatically. This has has fueled an increase in human genetics research. At the same time, with the recent advent of technologies in processing large data sets, lot of these technologies are proving valuable and efficient in analyzing these huge datasets. In this paper we use some of these technologies to analyze genetic sequencing data of 1000 Genomes Project,produce and evaluate a framework to process the sequencing data thereof and look into structural variations with respect to population groups.

\end{abstract}



%
\IEEEpeerreviewmaketitle

\section{Introduction}
\par{Projects like 1000 Genomes have propelled Next-Generation Sequencing (NGS), and have made available troves of mapped genome data for research. An increasing number of similar projects have adopted the same strategy and also made available data in the open domain. At the same time, the popularity of new frameworks like Hadoop\cite{white2009hadoop} MapReduce\cite{dean2008mapreduce}, Spark\cite{zaharia2010spark} and Storm\cite{leibiusky2012getting} have eased the difficulty which comes with managing and processing the mapped genome data which could easily go beyond terabytes in size. }
\par{Our scope is first delimited by 1000 Genomes which has cataloged all the human chromosomes for sample populations from around the world. We look at Chromosome 20 and filter out fragments that exhibit a specific type of structural variation in which the sample has had a deletion with respect to the reference genome. In order to evaluate our method's scaling we run on different-sized subsets of the data and using different configurations of system resources. We investigate frequency, locations and average length of such structural variation, and how these differ across geographical regions. We delimit our study to alignments that exhibit the structural variation in which the sample fragment is mapped to a fragment of the reference genome that is greater than 1000 base pairs in length, as opposed to the sample fragment length of 500, and analyze the following:

\begin{itemize}
\item Sample size (number of individuals) in  different   population groups.
\item Frequency of the specified structural variation in each of the populations, and sub populations. We also investigate this with respect to gender.
\item In which locations in the reference genome do they occur among the populations from the different geographical regions?
\item What is the average length of structurally variant fragments in the reference genome?
\end{itemize}


 



%
%


\section{Background}
\subsection{The 1000 Genomes Project}
The 1000 Genomes project was initiated by the European Bioinformatics Institute to establish a comprehensive and detailed map of human genome variations aiming to spawn studies to find disease-causing genes\cite{10002012integrated}. The project has mapped data and the variant genotypes for 1000+ individuals in 26 populations spread across 5 the super population groups of West African, European, American, and East- and South Asian. 


\subsection{Data}

The alignment data released by the 1000 Genomes project that this study is concerned with is made available in the BAM (Binary
Alignment/Map) format. This is the binary version of the SAM (Sequence Alignment/Map) format used by SAMtools software \cite{li2009sequence} and is a generic format for storing large nucleotide sequence alignments. With each BAM file there are two additional associated files containing indexing and statistics. For the purposes of our study the BAM file itself is sufficient and, the additional files were discarded. There is one BAM file for each individual and the alignment data of chromosome 20 contains data from 2535 individuals, resulting in a data set of size roughly 1 TB.For comparison, the entire dataset made available by 1000 genome project comprises of 43 TB of sequence data in FASTQ format, and 56 TB of alignment data in BAM format.

\hfill \linebreak


\subsection{Processing Large Data Sets}
One of the first models to process large data sets was released by Google and called MapReduce. This involved processing large data sets by slicing the source via mappers and consolidating the output by reducers. A popular implementation of the programming model is Apache's Hadoop.  This also gave the community a distributed file system HDFS which provided a fault-tolerant data store. Although it helped alleviate the processing and storing challenge, the immense size of data led the insatiable desire to learn and understand it more. Spark, a technology built atop Resilient Distributed Datasets\cite{zaharia2012resilient} gave a more of an end-to-end solution which addresses everything from loading of data to graphing the analyzed outcomes.

\hfill \linebreak
\hfill

\section{Related Work}

\textit{Genome Analysis Toolkit (GATK)} proposed by \cite{mckenna2010genome} emphasizes the importance of the MapReduce framework in churning next-generation sequenced data. The paper outlines the gap between the collected data and analysis outcomes, the collection size outweighing the capabilities for analysis. Contrary to McKenna et.al. we operated directly on the SAM\cite{sam} file from the object store, and parsed a readable TXT file instead of binary BAM file. Another work\cite{depristo2011framework} follows the McKenna et. al. study and adds on discovery of variants by machine learning. We carry out research in the context of Chromosome 20, and analyze variations across geographic regions.


\hfill \linebreak
\hfill

%
%

\section{Experimental Framework and Methods}

The main framework in which the experiment was executed is SMOG Cloud. This is an OpenStack based Infrastructure-as-a-Service (IaaS) resource provided by UPPMAX (Uppsala Multidisciplinary Center for Advanced Computational Science). 

\subsection{Openstack}
\par{OpenStack is a cloud computing software platform that provides processing, storage, and networking resources. It comprises several components in its architecture, each managing an aspect of the service it provides\cite{jackson2012openstack}. The relevant components for this project are Nova, which provides and manages the network of virtual machines that were used for computation and Swift object store, in which the raw data was initially stored.}

Two OpenStack instances of different configurations are used in our experiments

\begin{itemize}
\item Configuration 1: \texttt{m1.xlarge - 4 VCPUs, 160GB disk and 8GB RAM} 
\item Configuration 2: 3 instances combination Master-Slave with \texttt{m1.xlarge - 4 VCPUs, 160GB disk and 8GB RAM}  

\end{itemize}

\subsection{Apache Hadoop}

\par{The distributed processing framework provided by Apache Hadoop was used to analyze the alignment data. Hadoop implements the MapReduce programming model that allows distributed processing of large data sets across clusters of computers and includes its own distributed file system HDFS \cite{shvachko2010hadoop}. The particular tool used was Hive \cite{thusoo2009hive}, a system built on top of Hadoop that provides interactive querying using a SQL-like language and manages the conversion to MapReduce jobs transparently.}

\subsection{SAMtools}

SAMtools is a collection of software used to create and process DNA alignment data in the SAM/BAM format \cite{Li:2009:SAF:1613280.1613286}. This project utilized SAMtools to convert the given BAM files to the SAM format.
\hfill \linebreak

\section{Experiments}

\subsection{Methods}

\par{We approached the analysis by dividing the whole process into three subtasks, the first of which is pre-processing of data to make it available for use in Hive. The second task is the initial filtering in which the entire data set is scanned to find structurally variant alignments. The third step is analysis of the data extracted in the previous step to answer the specific questions stated. }

\par{The pre-processing of data consisted of reading the BAM files directly from the Swift object store, placing them on the local file system (FS), and then using SAMtools to do the conversion to SAM. As the Swift object store exposes the data objects via REST endpoints, a Swift Python command was modified to allow custom bulk-download of the files to FS. SAMtools was then used in-line to convert the files from BAM to SAM format.}

\par{The conversion to the text-format SAM was done to make it a format easily accessible to Hadoop MapReduce\cite{dean2008mapreduce}. The SAM files were then loaded into HDFS. These tasks were scheduled manually in the initial runs, and were later automated.}

\par{Once the SAM files were in HDFS, Hive was used for all subsequent analysis. Task 2 was the initial filter which extracted the subset of reads from chromosome 20 whose fragment length indicated by the alignment exceeded 1000 bp. These were then further analyzed through interactive querying in task 3 and the results either stored in Hive tables on HDFS for further processing, or written to CSV files for plotting.}

\par{HiveQL was used to further produce various relations to sub populations, and sub populations with respect to variants. Ancillary files which contained the pedigree of the individuals studied in the 1000 Genome project was loaded as Hive tables to assist with the ethnicity and gender aspect of the study. }

\par{The analysis of locations of structural variation was done by extracting the starting position in the reference genome of each alignment that exhibited the specified structural variation. This was subsequently grouped by geographical region.}

\par{The analysis was performed on two sets of the alignment data. Data set 1 consisted of 105 individuals from 5 super geographical regions and data set 2 consisted of the entire set of 2535 individuals. }

\par{R\cite{team2000r} and Python\cite{chun2001core} was used to create the visualizations. }


\section{Results}

The results of the analysis of structural variation in chromosome 20 for the entire data set of 2535 individuals are shown below.

\subsection{Sample Size per Population Group}

\includegraphics[scale=0.38]{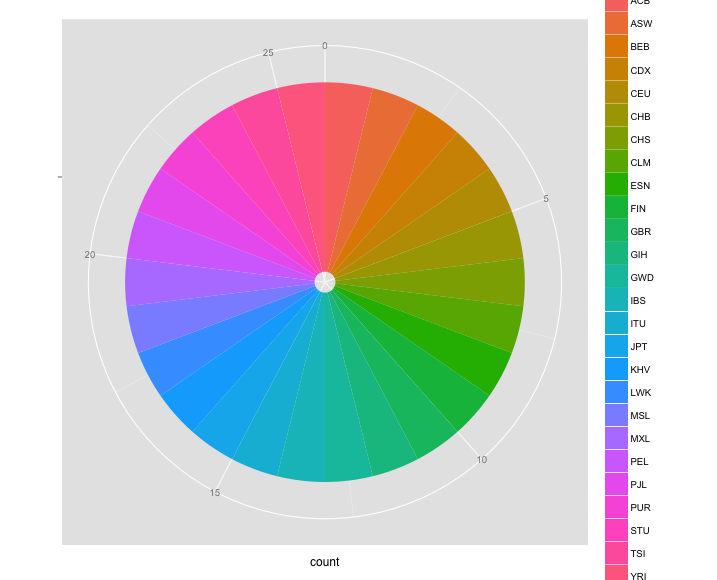}

\subsection{Frequency of Structural Variation per Population Group}

\includegraphics[scale=0.38]{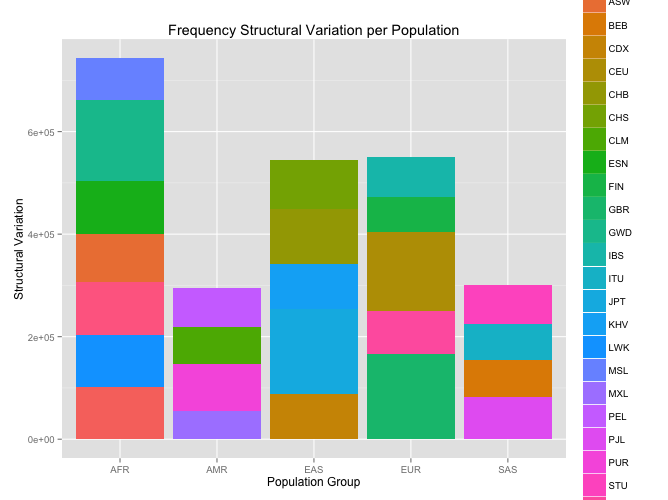}
\centerline{}

\subsection{Frequency of Structural Variation per Population Supergroup}

\includegraphics[scale=0.35]{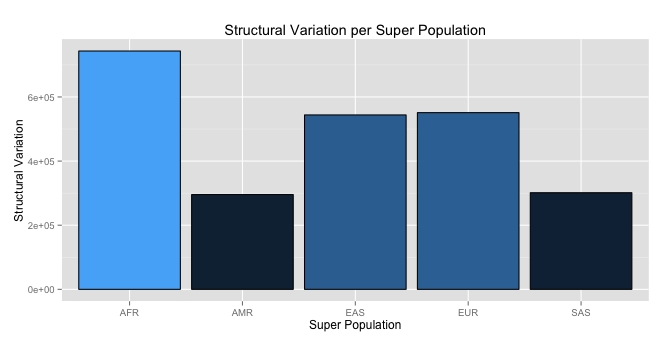}
\centerline{}
\hfill \linebreak

\subsection{Frequency of Structural Variation per Population Group and Gender}

\includegraphics[scale=0.45]{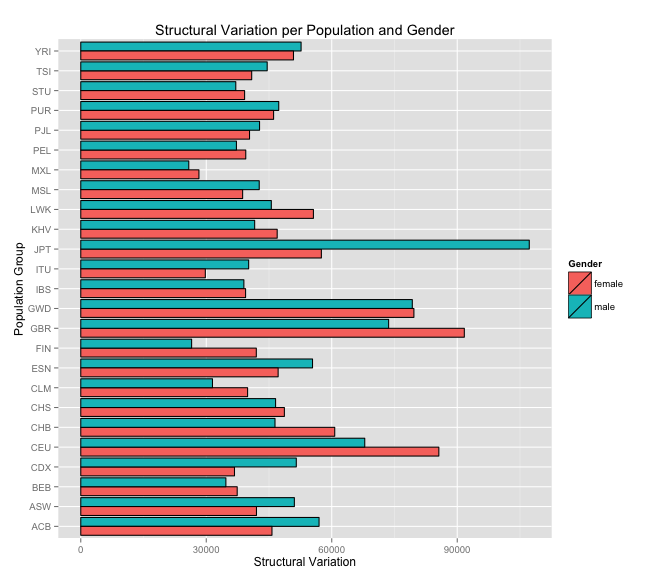}
\hfill \linebreak

\subsection{Frequency of Structural Variation per Population Supergroup and Gender}

\includegraphics[scale=0.38]{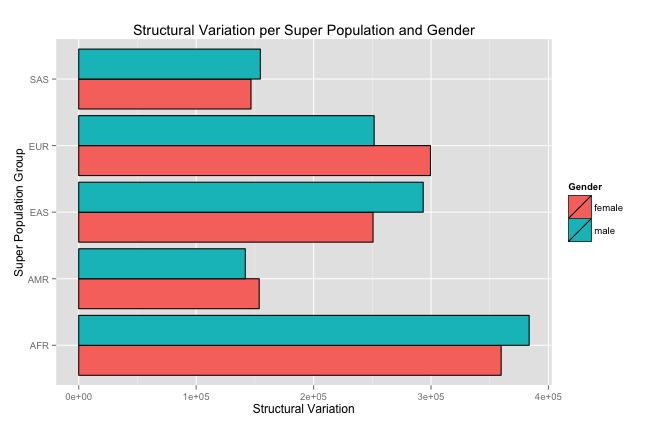}
\hfill \linebreak

\subsection{Location of Structural Variation in the Reference Genome per Population Group}

\includegraphics[scale=0.34]{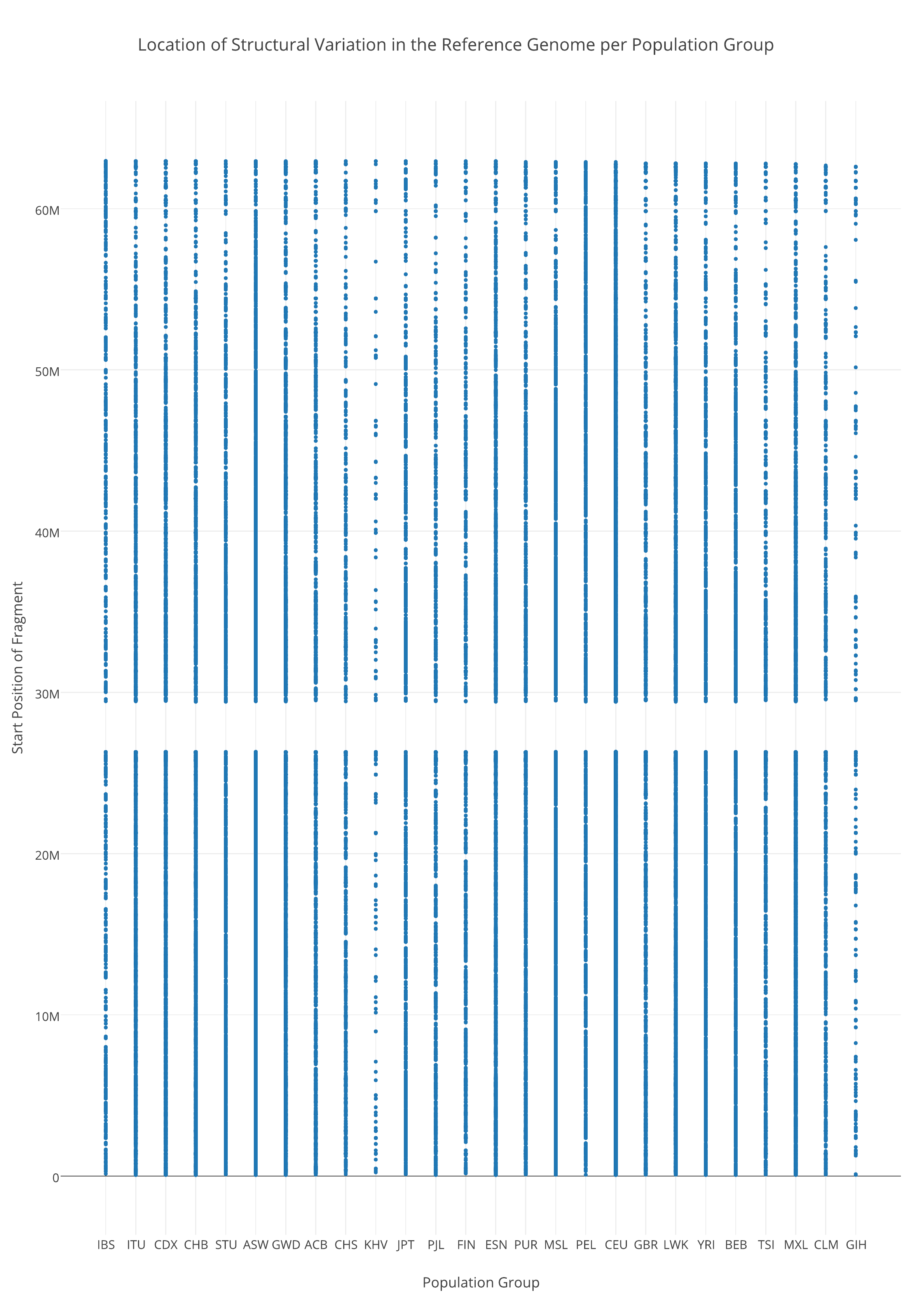}
\centerline{ ...}
\hfill \linebreak

\subsection{Average Length of Alignment per Population}

\includegraphics[scale=0.40]{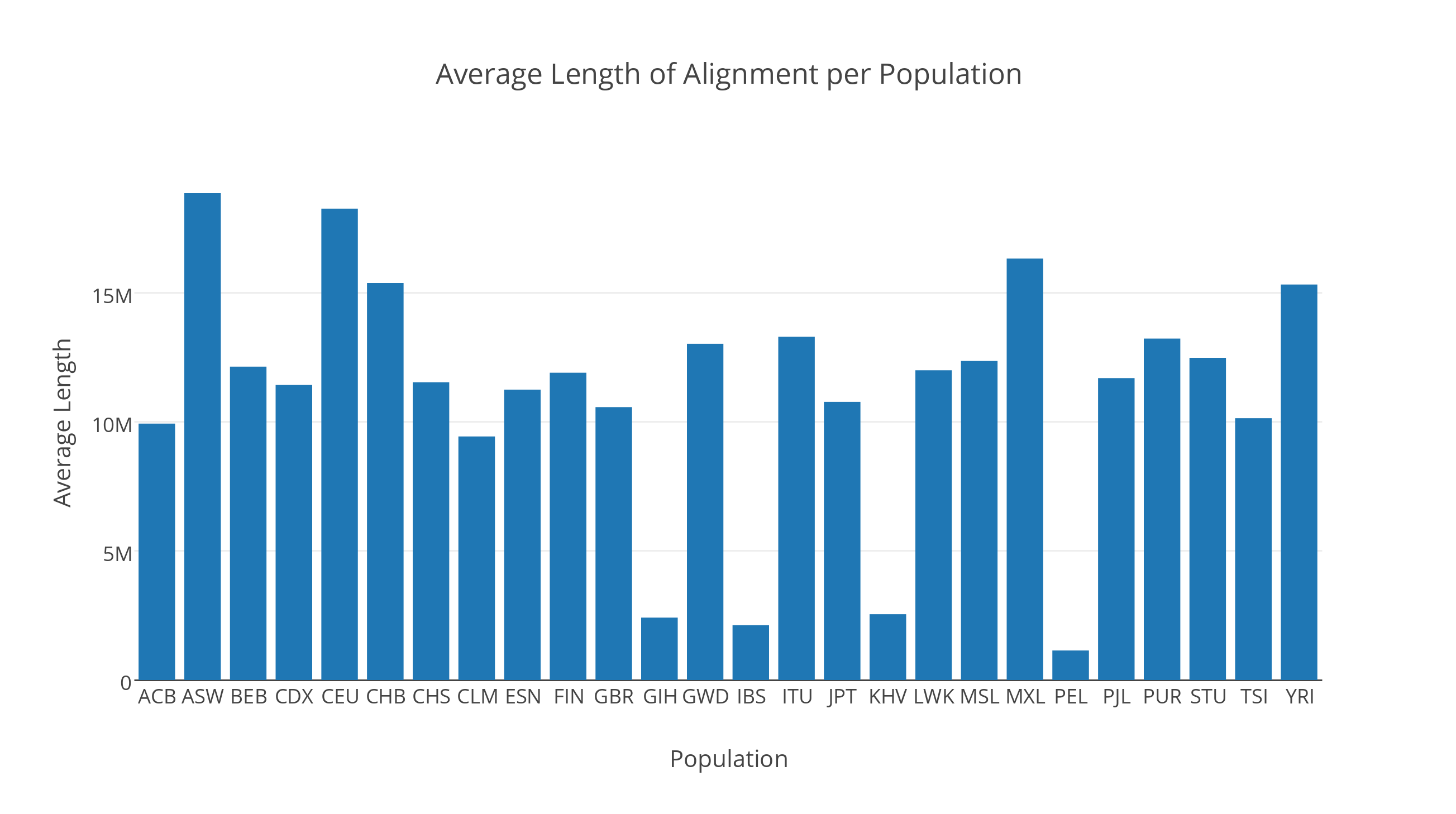}
\centerline{}
\hfill \linebreak

\section{Discussion}

\subsection{Analysis of Structural Variation}

\par{The results of the analysis of data set 1 show some interesting patterns. They indicate differences in both frequency, position and average length of the structurally variant genome fragments.} 

\par{The African ancestry superpopulation (AFR) stands out as the one with the highest number of this particular variation. The fact that the distribution between the sub populations within AFR is quite similar indicates that this tendency might be true in general for the entire superpopulation. This is not the case for the European ancestry (EUR) superpopulation which had the second highest frequency overall. In this case there is one population, Utah residents with Northern and Western European ancestry (CEU), that stands out with much higher frequency than the rest.}

\par{When considering the gender-separated frequency data it becomes clear that it is the women of the CEU population that exhibit high frequencies. Indeed, this graph shows that the women of this group are a clear outlier among all other population- and gender groups. However, the sample size considered in this study is for preliminary finding, the analysis ought to be run on the entire set of genome data for concreteness, the size of which is around 50x the size used in this study.} 

\par{The mapping of the starting positions of the fragments in the reference DNA show some general patterns. There seem to be less fragments mapped towards the end of the reference genome, but otherwise spread throughout the entire length. The area between 26000-29000 bp contains none at all because that corresponds to the centromeric region which is non sequenced.}

\par{The data of the average length of these alignments in the reference genome can be used to compare the average length of deletions in the different population groups. The results suggest that the populations with the lowest frequencies of deletions also exhibited lowest average length of the deletions.}

\subsection{Evaluation of the Computing Framework}

\par{The goal of the evaluation was to try both data sets on both configurations. The intention was to get an indication of how the processing times changed when data size was increased on the same configuration, as well as how the times were affected when using more computing resources on the same data set. However, due to lack of resources and time this was not completely realized and the scaling analysis had to be slightly modified.}

\par{Data set 1 was analyzed on both configurations, and the times taken to process each of the three tasks are illustrated in Table I.}

\begin{table}[h]
\begin{center}
\begin{tabular}{l*{6}{l}r}
\hline
Task  & Stage & Config 1 (Secs) & Config 2 (Secs) \\
  
\hline
 1 & Pre-processing of data & 3120  & 180000  \\
 2 & Initial Filtering & 555   &  60  \\
 3 & Subsequent Filtering & 18  &  - \\
\hline
\end{tabular}
\end{center}
\caption{Execution Times of Data Set 1 on the Different Configurations}
\end{table}

\par{The summary of the runs shows that the pre-processing of data is the main bottleneck. This is reasonable as this is the part that does the most amount of reading and writing to disk, and in particular since this is done sequentially file by file. While task 2 filters through the entire data set it is done using Hive which innately uses MapReduce in a distributed manner, and the results show that this is relatively quick. The subsequent queries to answer the specific questions are all very quick as they operate on a much reduced data set. Comparing the runtimes of the tasks of data set 1 on the different configurations shows that the tasks that use Hive were much faster on configuration 2, as expected. Task 2 was faster by a factor of around 9. Task 3 was not possible to run due to a malfunctioning of the machine and lack of time, but it is reasonable to assume that it would scale similarly as task 2. The fact that task 1 took more time on configuration 2 was due to different implementations of the reading routine from Swift object store - In configuration 1, it was bulk read of 5 files per run i.e. one authentication token per run, and in configuration 2, it was reading the file through a driver file which consisted the entire set of file names i.e. requested one token for each fetch, hence, these cannot be compared. However, since task 1 was sequential and focused on I/O it was not merited on how it scaled on the different configurations.}

\par{Due to restrictions in disk size the three tasks had to be modified when trying data set 2 which is of size 1.1 TB. The pre-processing and initial filtering were combined into one task that was executed sequentially on each file. While this reduces the disk space required since the entire unfiltered set of SAM files is never stored on HDFS, it also resulted in a large task that took a very long time, around 96 hours on configuration 1. The main reason why the time did not increase linearly with respect to data size compared to the initial run on a smaller data set is probably that this time a MapReduce job needs to be set up, started, run and terminated once per file, as opposed to once in total once all the SAM files are in HDFS. The large increase in time shows that this overhead is very significant and indicates that it would not be much quicker when run on configuration 2. For this reason and lack of time the complete data set was only processed on configuration 1.}

\par{Because tasks 1 and 2 were modified when processing data set 2, the execution times for them when varying data size but not configuration are not comparable. The third task, however, remained the same and its times are possible to compare for configuration 1. This is because it consisted of queries being run on the Hive table consisting of the already filtered-out alignments. This took on average 18s on the smaller data set, and 45s on the larger data set. In the first case the table contained 121476 entries, and in the second 2516951 entries. So increasing data size by a factor of 20 resulted in an increase in time by a factor of roughly 2.5 on configuration 1. This indicates that Hive queries scale very well when increasing data size, and further suggests that a large portion of the time involved in MapReduce tasks is the setup of the job.}

\section{Conclusions and Future Work}

\par{In conclusion, we developed an application which accessed and loaded the alignment data for chromosome 20 and analyzed it using Hive. The results provided some insights regarding patterns of structural variation inf the different population and gender groups. An evaluation of the scalability of the solution was also performed.}

\par{The results that we did manage do gain on scaling indicate that the pre-processing is the main problem. This scaled poorly because of the sequential manner it was performed in. The amount of space on disk was also a limitation. Finally, the results show that Hive itself scales well. The main issue is to get the data on HDFS in a format that Hive can use, once that is done the querying itself is fast and scalable up to the largest data set tried.}

\par{As the largest bottleneck of the tasks was the pre-processing of data, future work would be focused on this area. There are two obvious improvements, the first of which would be eliminating the middle step of writing to the local file system between reading from Swift and writing to HDFS. The second would be to handle the process of converting from BAM files on Swift to SAM files on HDFS in a distributed fashion instead of one file at a time. Another strategy that could improve scaling could be to use BAM files on HDFS directly. There exists libraries for handling this format in Hadoop MapReduce as well as Hive already. Lastly, another approach to look into is to use Spark to directly read from the Swift data store and create Resilient Distributed Datasets which can be partitioned and processed efficiently in a distributed manner in the Spark framework.}


\section*{Acknowledgments}
 
Authors want to thank Andreas Hellander,
Associate Professor, Uppsala University for providing feedback along the experiments.




%


\bibliographystyle{IEEEtran}
\bibliography{ref} 

\end{document}